\documentclass[12pt]{article}
\pdfoutput=1
\usepackage{amsmath}
\usepackage{amsfonts}
\usepackage{setspace}
\usepackage{subcaption}
\usepackage{enumerate}
\usepackage{graphicx}
\usepackage[hidelinks]{hyperref} 
\usepackage[utf8]{inputenc}
\newcommand{\gsim}{\lower.7ex\hbox{$\;\stackrel{\textstyle>}{\sim}\;$}}
\newcommand{\lsim}{\lower.7ex\hbox{$\;\stackrel{\textstyle<}{\sim}\;$}}
\def\O{{\mathcal O}}

\newcommand{\be}{\begin{equation}}
\newcommand{\ee}{\end{equation}}
\newcommand{\bea}{\begin{eqnarray}}
\newcommand{\eea}{\end{eqnarray}}

\newcommand{\comment}[1]{}

\newcommand{\Cint}{C\kern-1em\int}

\def\d{\partial}

\def\O{\mathcal{O}}

\def\Im{{\rm Im ~}}
\def\Re{{\rm Re~}}

\def\Z{{\mathcal Z}}
\begin{document}
\vspace*{-1. cm}
\begin{center}
{\bf \Large A 2-Replica Wormhole}
\vskip 1cm
{{\bf Mehrdad Mirbabayi} }
\vskip 0.5cm
{\normalsize {\em International Centre for Theoretical Physics, Trieste, Italy}}
\vskip 0.2cm
{\normalsize {\em Stanford Institute for Theoretical Physics, Stanford University,\\ Stanford, CA 94305, USA}}
\end{center}
\vspace{.8cm}
{\noindent \textbf{Abstract:} 
Replica geometries are not rigid when gravity is dynamical. We numerically construct $1<n\leq 2$ replica saddles in $2d$ gravity coupled to a CFT and compare the resulting Renyi entropies with the field theory result.
\vspace{0.3cm}
\vspace{-1cm}
\vskip 1cm
\section{Introduction}
The von Neumann entropy can be obtained by formally calculating the partition function of $n$ copies of the system, $Z_n$, as an analytic function of $n$, properly normalizing to get the Renyi entropy,
\be\label{Sn}
S_n =\frac{1}{1-n} \log(Z_n/Z_1^n),
\ee
and taking the limit $n\to 1$. This {\em replica} method has played a central role in understanding how and why we can use semiclassical gravity to calculate entropy \cite{Lewkowycz,Faulkner,Rangamani,Dong,Yang,Almheiri}. 

There is a novelty when the method is applied to gravity. The replica geometry on which $Z_n$ is calculated is integrated over in the gravitational path integral. For instance, the number, location, and shape of the entangling surfaces have to be determined dynamically. For some interesting consequences of this fact see \cite{Hartman,Penington,Marolf,Zhao}. These variables are the relevant gravitational degrees of freedom when $n\approx 1$. The goal here is to give an example with finite $n-1$, in which other geometric degrees of freedom have a noticeable effect on the Renyi entropy.

\section{$2d$ setup and the results}
The problem we study is formulated in \cite{Almheiri}, in a model that consists of an AdS$_2$ black hole coupled to a thermal CFT (see \cite{Mathur,Mahajan} for earlier incarnations). The entropy is that of a single interval $[-a,b]$, partly in the gravitational region (see figure \ref{fig:setup}). Generically there is a nontrivial diffeomorphism $\theta(\tau)$ at the boundary of the two regions. To calculate $Z_n$, via saddle point approximation to the gravity path integral, we need to find $a$ and $\theta(\tau)$ that solve
\begin{figure}[t]
\centering
\includegraphics[scale =0.9]{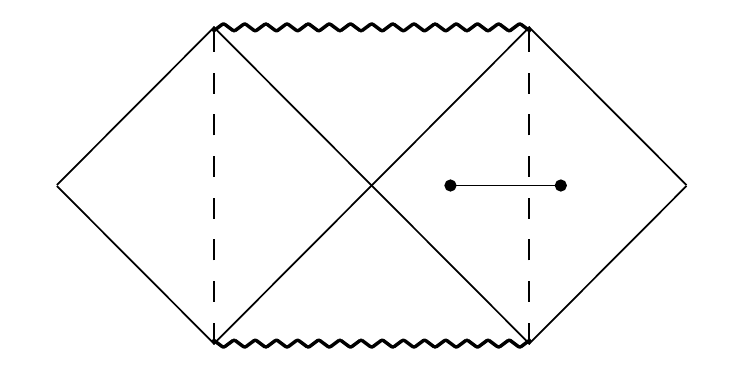} 
~~~~~~~~~~~\includegraphics[scale =1.6]{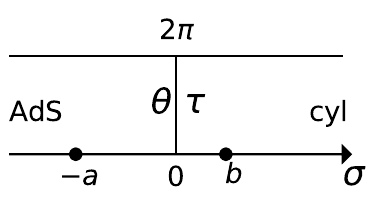}~~~~~~ 
\caption{\small{Left: The Penrose diagram of AdS$_2$ black hole connected along the UV boundary (dashed lines) to the non-gravitational region. CFT fields freely propagate through this boundary and are in thermal equilibrium with the black hole. We are calculating the Renyi entropies of an interval anchored on one side in the flat region. Right: The Euclidean geometry that prepares the state. Since the boundary curve between AdS and half-cylinder fluctuates, mapping it to $\sigma =0$ generically requires a nontrivial diffeomorphism $\theta(\tau)$. $b$ and $-a$ are the entangling points.}}
\label{fig:setup}
\end{figure}
\be\label{replica}
\d_\tau \left[\{e^{i\theta},\tau\}+\frac{1}{2}(1-\frac{1}{n^2}) R(\theta(\tau))\right] = 
i \kappa \left[-\frac{1}{2}\left(1-\frac{1}{n^2}\right)\left(\frac{(1+ F') z (z_b-z_a)}{(z-z_a)(z-z_b)}\right)^2-\{e^{y+F},y\}\right]+c.c.
\ee
where $\{f,x\}=(f''/f')'-\frac{1}{2} (f''/f')^2$, 
\be
R(\theta) = -\frac{(1-A^2)^2 \theta'^2}{|1- A e^{i\theta}|^4},\qquad A \equiv e^{-a},
\ee
and $y = i\tau$, $z = e^{y + F(y)}$, $z_a = e^{-a + G(-a)}$, $z_b = e^{b+F(b)}$. $F$ and $G$ are holomorphic functions on complex half-strips:
\be\label{frame}\begin{split}
F(y) &= \sum_{n\geq 0} f_n e^{- n y},\qquad \Re y >0,\quad \Im y \sim \Im y + 2\pi\\[10pt]
G(x) &= \sum_{n>0} g_n e^{n x},\qquad \Re x <0,\quad \Im x \sim \Im x + 2\pi.
\end{split}
\ee
They are fixed, given $\theta(\tau)$, by the condition
\be\label{weld}
i\tau + F(i\tau) = i\theta(\tau) + G(i\theta(\tau)).
\ee
This is called the conformal welding problem \cite{Sharon}: $e^{x+G(x)}$ and $e^{y+F(y)}$ are conformal maps from the two sides of the Euclidean strip with the gluing function $\theta(\tau)$ to the interior and the exterior of a curve in the complex $z$-plane. The $SL(2,C)$ symmetry has been fixed by the choice \eqref{frame}.

$b$ is a parameter of the solution. It is unambiguously specified in the non-gravitational region. $\kappa$ is an effective coupling constant. $a$ is a redundant variable. It can be mapped by AdS isometries to any point, at the cost of changing $\theta$ by an $SL(2,R)$ transformation. We fix $a$ by
\be\label{a}
2\kappa \sinh(a) \sinh\frac{a-b}{2} = \sinh \frac{a+b}{2}.
\ee
With this choice $\lim_{n\to 1} \theta(\tau) = \tau$. 

Our numerical results for the Renyi entropy as well as some diagnostics of the change in the geometry as $n$ varies from $1$ to $2$ are shown in figure \ref{fig:result}.
\begin{figure}[t]
\centering
\includegraphics[scale =0.9]{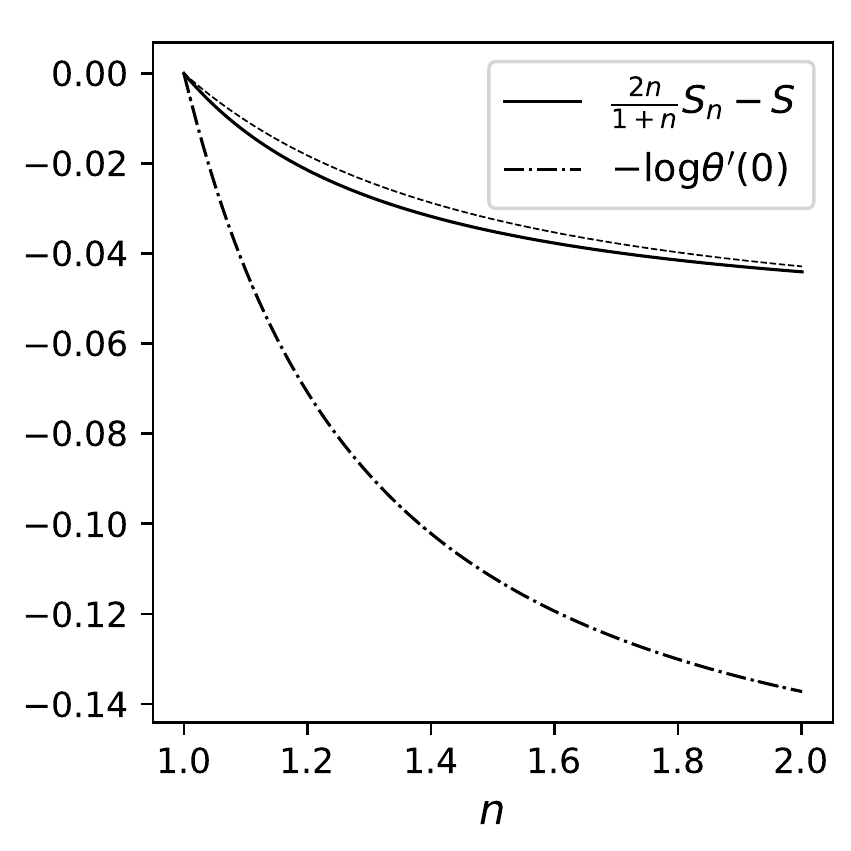} 
\includegraphics[scale =0.9]{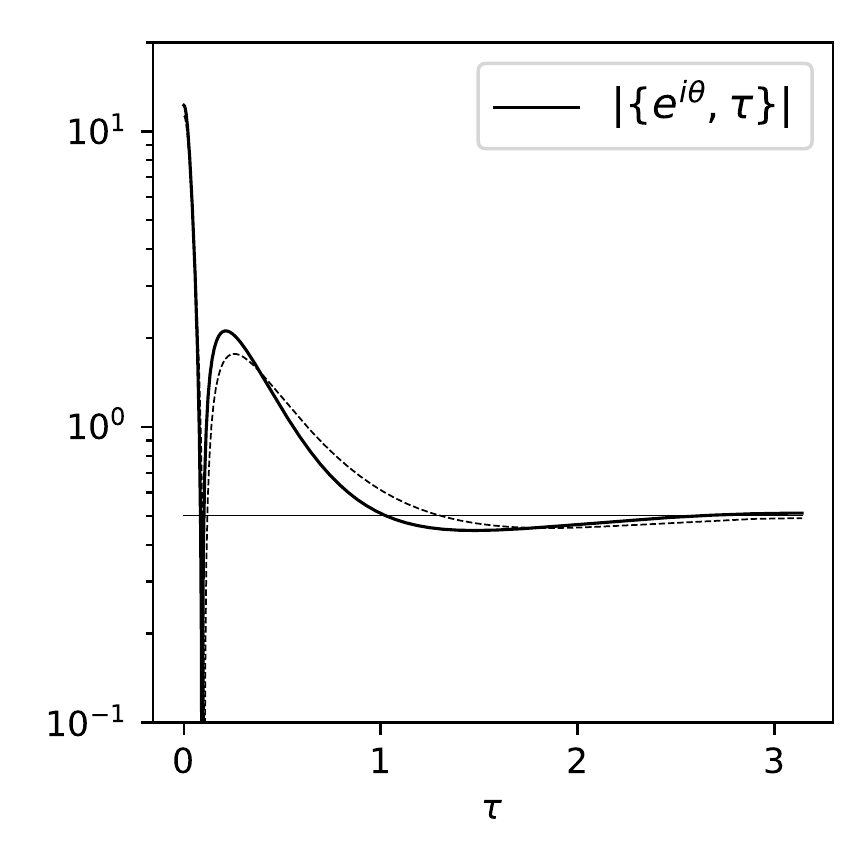} ~~~~~~
\caption{\small{Left: Solid line shows the Renyi entropy, multiplied by $2n/(1+n)$ to make the cutoff-dependent contribution constant, and shifted by the von Neumann entropy. This combination vanishes for a fixed interval in a CFT. The dashed line shows the result obtained by linearizing the equations in $\theta-\tau$. The dot-dashed line shows the change in the proper distance between $x = -a$ and $x=0$. Right: The Schwarzian of the AdS boundary-time at $n=2$, as a measure of how the boundary has deformed. This is $1/2$ for an undeformed boundary (the horizontal line). The dashed line shows the linear approximation. The parameters are $b=0.05$, $\kappa =1$.}}
\label{fig:result}
\end{figure}
\section{Partition function}
Given $F$ and $G$, the standard expression for the CFT partition function can be obtained, up to a coefficient $\Z$, from the correlation function of twist operators with conformal dimensions $\Delta =\bar\Delta = \frac{c}{24}(1-\frac{1}{n^2})$ \cite{Cardy}:
\be\label{Ztw}
Z_n^{\rm mat} = \mathcal{Z}\ \left( (1+F'(b))(1+G'(-a))\sinh(a) \frac{z_a z_b}{(z_b-z_a)^2}\right)^{c\left(n- \frac{1}{n}\right)/12}.
\ee
In quantum field theory $\Z$ is uninteresting because it does not depend on $a$ and $b$. In our case, it does since the conformal maps depend on $\theta$, which is ultimately related to $b$. This dependence enters the partition function via the conformal anomaly:
\be\label{anom}
Z[g= e^{2\omega} \hat g] = \exp\left(-\frac{c}{24 \pi} \int \sqrt{\hat g} \left[ \omega \hat R + (\hat\nabla \omega)^2\right]\right)Z[\hat g].
\ee
It is convenient to do this in two steps. First, scale away the AdS warp factor $1/\sinh^2 \sigma$. This would give an uninteresting contribution because it is independent of $\theta$. Then the line element is simply $ds^2 = dx d\bar x$ on the left side of the strip and $ds^2= dy d \bar y$ on the right side. Therefore
\be
2\omega_{\rm out} = \log \left|\frac{\d y}{\d z}\right|^2 , \qquad 2\omega_{\rm in } = \log \left|\frac{\d x}{\d z}\right|^2.
\ee
It is now necessary to add the following boundary term to \eqref{anom}
\be\label{K}
\log \Z_{\rm boundary} = - \frac{c}{12\pi}\oint\omega \hat K
\ee
since in general $\omega_{\rm in}(\theta(\tau),0)\neq \omega_{\rm out}(\tau,0)$. The extrinsic curvature of the boundary is given by
\be
\hat K(\tau) = \frac{\Re (z'(i\tau) \bar z''(-i\tau))}{|z'(i\tau)|^3}.
\ee
Transforming the integration domain back to the strip and discarding the constant part, we obtain
\be
\frac{1}{n}\log \Z = \frac{c}{48 \pi}\int_0^{2\pi} d\tau \left[2 (F +\bar F) + (\bar F+ \log(1+ \bar F'))\left(F' + \frac{F''}{1+F'}\right)\right]
-(\tau \to \theta, F\to G),
\ee
where $F$ and its derivatives, are holomorphic functions of $i\tau$. The difference between this expression and \eqref{K} is a contribution from $(\hat\nabla \omega)^2$ in \eqref{anom} that cancels half of the nonlinear piece. We have also used the fact that $F$ contains only non-positive frequencies to eliminate some terms. 

One can check that varying \eqref{Ztw} with respect to $\theta$ gives the correct stress-energy tensor $T_{\tau\sigma}$, which is the source term on the right-hand side of \eqref{replica}. The left-hand side follows from varying the gravitational action \cite{Almheiri}
\be
- I_{\rm grav} = S_0 + \frac{\phi_r}{2\pi}n \int_0^{2\pi} d\tau \left[\{e^{i\theta},\tau\} +\frac{1}{2}\left(1-\frac{1}{n^2}\right) R(\theta)\right],
\ee
where $S_0$ is the extremal entropy, and $\phi_r$ controls the asymptotic growth of dilaton in AdS. In terms of $\phi_r$, $c$ and inverse temperature $\beta$, which is always set to $2\pi$, $\kappa =\beta c/24\pi \phi_r$. 

To obtain the plot in figure \ref{fig:result}, we used the saddle-point approximation to $Z_n/Z_1^n$:
\be
\log \frac{Z_n}{Z_1^n} \approx - I_{\rm grav}- n S_0 + \log Z_n^{\rm mat}.
\ee
\section{Integration}
Equation \eqref{replica} is in general both nonlinear and nonlocal in $\theta(\tau)$. However, in various limits the deformation of the boundary curve is small, and by an appropriate choice of $a$ one can perturb in
\be
H(\tau) = \theta(\tau)- \tau.
\ee
The first two limits are discussed in \cite{Almheiri}. One is $n\to 1$, in which case we approach the background solution. The choice of $SL(2,R)$ frame that makes $H\to 0$ is \eqref{a}. The second limit is $\kappa\ll 1$. In this case the energy-momentum flux through the AdS boundary is small. Hence the boundary curve remains close to a circle and $z_a$ would sit close to the center (i.e. $a\to \infty$). The third limit is $b\gg 1$, and by \eqref{a} also $a\simeq b \gg 1$. In this case we recover an approximate shift symmetry in $\tau$. If the symmetry were exact, $H$ would be zero because it is an odd function of $\tau$. So $H\ll 1$ when $b\gg 1$. Finally, if $\kappa \gg 1$, we should find a configuration that suppresses the right side of \eqref{replica}, which is $T_{\tau\sigma}$. This flux would exactly vanish if the boundary curve were the circle $|z|=1$ and $z_a$ were at the image of $z_b$ (i.e. $z_a = 1/z_b$). Since \eqref{a} implies $a- b\ll 1$ in this limit, $H$ should also be small to approach this configuration. 

Away from these limits, even though it is not straightforward to solve \eqref{replica}, for a given $n,b$ and $H$ (and hence also given $F$ and $G$) it is relatively easy to find $dH/dn$. Since $H=0$ at $n= 1$, we can continuously vary $n$ from $1$ and integrate $dH/dn$. The plots in figure \ref{fig:result} have been obtained by taking steps $\Delta n =0.01$. Since the explicit $n$-dependence in \eqref{replica} is of the form $1- 1/n^2$, it doesn't take long to converge to the asymptotic $n\to \infty$ solution. 

Since the most computationally expensive step is to find the welding maps, we will discuss it in some detail. For a given $H$, the Fourier modes of $F$ and $G$ can be solved from
\be\label{weld1}
\sum_{n\geq 0} f_n e^{-in\tau} - \sum_{n>0} g_n e^{in (\tau+H(\tau))} = i H(\tau).
\ee
We truncate the sum over $n$ at $N$ and take $N_\tau$ samples of $\tau$ in the range $[0,\pi]$. To get a reliable answer, we choose $N_\tau\gg N$ and find the pseudo-inverse of the resulting matrix of coefficients $C$ using python.scipy.pinv. $N$ has to be larger for smaller $b$. Our plots are obtained with $N=400$. This algorithm seems conceptually easier than solving the $2d$ boundary value problem discussed in \cite{Sharon}. However, it might not be faster. A brute-force method to solve $C x = p$ for a non-square matrix $C$ is to multiply both sides by $C^T$ and apply the conjugate gradient method to solve $(C^T C) x = C^T p$ which is $\O(N^2)$.\footnote{See \cite{CG} for an implementation, with the caveat that our matrix is not the Laplace operator.} We found this to be only slightly slower than scipy.pinv. 

For our purpose, namely to find $\frac{d H}{dn}(H,F,G,n,b)$, we need to solve not for $f_n$ and $g_n$ but rather $\d f_n/\d H(\tau)$ and $\d g_n/ \d H(\tau)$ as functions of $H,F,G$, i.e. to invert
\be
\sum_{n\geq 0} \frac{\d f_n}{\d H(\tau)} e^{-i n\tau}-\sum_{n>0}\frac{\d g_n}{\d H(\tau)}e^{i n (\tau + H(\tau))} = i + i\sum_{n>0} n g_n e^{i n(\tau+H(\tau))}.
\ee
The same methods can be applied to find these matrices. This takes $\O(N^3)$. We will also need $F_H(y,\tau)\equiv \d F(y)/\d H(\tau_2)$ and $G_H(x,\tau)\equiv \d G(x)/\d H(\tau)$ obtained by Laplace or Fourier transforming the result. We can now derive a linear equation $ L \frac{dH}{dn} = p$, where
\be\label{L}\begin{split}
L =&\frac{1}{1+ H'}\d_\tau^4 - \frac{ H'''' }{(1+ H')^2}\d_\tau
-4\frac{1}{(1+H')^2}(H''\d_\tau^3+ H'''\d_\tau^2)+8 \frac{H'' H''' }{(1+H')^3} \d_\tau \\[10 pt]
&+9 \frac{H''^2 }{(1+H')^3}\d_\tau^2-9 \frac{H''^3 }{(1+H')^4}\d_\tau
+(1+H') \d_\tau^2+ H'' \d_\tau\\[10pt]
&+\frac{1}{2}(1-\frac{1}{n^2})\left(
 \frac{R'_1+3 R'_2}{1+H'}\d_\tau+ \frac{R'_1}{H''} \d_\tau^2
-2 \frac{ A\sin\theta }{1+ A^2-2 A\cos\theta}(2 R'_1+3 R'_2)
+R'_2 \cot \theta \right)\\[10pt]
&+2 \kappa \Im \Big(
-\frac{F_H'''}{1+ F'} + \frac{F''' F_H'}{(1+F')^2}+3\frac{F'' F_H''}{(1+ F')^2}
-3 \frac{ F''^2 F_H'}{(1+F')^3}+(1+F') F_H'\\[10 pt]
&~~~~~~~~~~~~~+2 T_n \frac{F_H'}{1+ F'}+ 2 T_n (1-\frac{z}{z-z_a}- \frac{z}{z-z_b}) F_H\\[10pt]
&~~~~~~~~~~~~~+2 T_n(\frac{z_b}{z_b-z_a}+\frac{z_b}{z-z_b}) F_H(b)
+  2 T_n ( \frac{z_a}{z_a -z_b} +\frac{z_a}{z-z_a}) G_H(-a)\Big)\\[10 pt]
&p =\frac{1}{n^3}\left[-\d_\tau R+2\kappa \Im \Big(\frac{ z^2(z_b- z_a)^2(1+F')^2}{(z-z_a)^2(z-z_b)^2}\Big)\right]
,\end{split}
\ee
with
\be
R'_1 = - 2 \frac{(1-A^2)^2 \theta' \theta''}{|1- A e^{i\theta}|^4},\quad 
R'_2 =  4 \frac{ A (1-A^2)^2 \theta'^3 \sin\theta}{|1-A e^{i\theta}|^6}
\ee
and 
\be
T_n =
-\frac{1}{2}\left(1-\frac{1}{n^2}\right)\left(\frac{(1+ F') z (z_b-z_a)}{(z-z_a)(z-z_b)}\right)^2.
\ee
The arguments of functions are suppressed unless there is an ambiguity. The terms on the l.h.s. of \eqref{replica} give local contributions to $L$, i.e. close to diagonal in the discrete problem, while those coming from the r.h.s. are nonlocal. As an example, $F_H'''/(1+F') = \frac{1}{(1+F'(i\tau_1))}\d_{i\tau_1}^3 F_H(i\tau_1,\tau_2)$. Finally, in practice it is convenient to work with the Fourier modes $h_n$ defined by
\be
H(\tau) = \sum_{n} h_n e^{i n\tau},\qquad h_{-n} = h_{n}^*,\qquad \Re h_n = 0.
\ee
The corresponding matrices are obtained by applying FFT to the ones written for $H$. We integrate the system forward by inverting $L$ and using $F_H$ and $G_H$ to update $F$ and $G$ at each step. The linear approximations in figure \ref{fig:result} are obtained by setting $f_n = g_n = \Im h_n$, and linearizing \eqref{replica} in $h_n$. This can be inverted at any replica number.

\section*{Acknowledgments}
We thank Raghu Mahajan, Douglas Stanford, Steve Shenker, and Zhenbin Yang for useful discussions. This work was partially supported by the Simons Foundation Origins of the Universe program (Modern Inflationary Cosmology collaboration).
\appendix
\bibliography{bibrep}

\providecommand{\href}[2]{#2}\begingroup\raggedright\begin{thebibliography}{10}

\bibitem{Lewkowycz}
A.~Lewkowycz and J.~Maldacena, ``{Generalized gravitational entropy},''
  \href{http://dx.doi.org/10.1007/JHEP08(2013)090}{{\em JHEP} {\bfseries 08}
  (2013) 090},
\href{http://arxiv.org/abs/1304.4926}{{\ttfamily arXiv:1304.4926 [hep-th]}}.

\bibitem{Faulkner}
T.~Faulkner, A.~Lewkowycz, and J.~Maldacena, ``{Quantum corrections to
  holographic entanglement entropy},''
  \href{http://dx.doi.org/10.1007/JHEP11(2013)074}{{\em JHEP} {\bfseries 11}
  (2013) 074},
\href{http://arxiv.org/abs/1307.2892}{{\ttfamily arXiv:1307.2892 [hep-th]}}.

\bibitem{Rangamani}
X.~Dong, A.~Lewkowycz, and M.~Rangamani, ``{Deriving covariant holographic
  entanglement},'' \href{http://dx.doi.org/10.1007/JHEP11(2016)028}{{\em JHEP}
  {\bfseries 11} (2016) 028},
\href{http://arxiv.org/abs/1607.07506}{{\ttfamily arXiv:1607.07506 [hep-th]}}.

\bibitem{Dong}
X.~Dong and A.~Lewkowycz, ``{Entropy, Extremality, Euclidean Variations, and
  the Equations of Motion},''
  \href{http://dx.doi.org/10.1007/JHEP01(2018)081}{{\em JHEP} {\bfseries 01}
  (2018) 081},
\href{http://arxiv.org/abs/1705.08453}{{\ttfamily arXiv:1705.08453 [hep-th]}}.

\bibitem{Yang}
G.~Penington, S.~H. Shenker, D.~Stanford, and Z.~Yang, ``{Replica wormholes and
  the black hole interior},''
\href{http://arxiv.org/abs/1911.11977}{{\ttfamily arXiv:1911.11977 [hep-th]}}.

\bibitem{Almheiri}
A.~Almheiri, T.~Hartman, J.~Maldacena, E.~Shaghoulian, and A.~Tajdini,
  ``{Replica Wormholes and the Entropy of Hawking Radiation},''
  \href{http://dx.doi.org/10.1007/JHEP05(2020)013}{{\em JHEP} {\bfseries 05}
  (2020) 013},
\href{http://arxiv.org/abs/1911.12333}{{\ttfamily arXiv:1911.12333 [hep-th]}}.

\bibitem{Hartman}
T.~Hartman and J.~Maldacena, ``{Time Evolution of Entanglement Entropy from
  Black Hole Interiors},''
  \href{http://dx.doi.org/10.1007/JHEP05(2013)014}{{\em JHEP} {\bfseries 05}
  (2013) 014},
\href{http://arxiv.org/abs/1303.1080}{{\ttfamily arXiv:1303.1080 [hep-th]}}.

\bibitem{Penington}
G.~Penington, ``{Entanglement Wedge Reconstruction and the Information
  Paradox},'' \href{http://arxiv.org/abs/1905.08255}{{\ttfamily
  arXiv:1905.08255 [hep-th]}}.

\bibitem{Marolf}
A.~Almheiri, N.~Engelhardt, D.~Marolf, and H.~Maxfield, ``{The entropy of bulk
  quantum fields and the entanglement wedge of an evaporating black hole},''
  \href{http://dx.doi.org/10.1007/JHEP12(2019)063}{{\em JHEP} {\bfseries 12}
  (2019) 063}, \href{http://arxiv.org/abs/1905.08762}{{\ttfamily
  arXiv:1905.08762 [hep-th]}}.

\bibitem{Zhao}
A.~Almheiri, R.~Mahajan, J.~Maldacena, and Y.~Zhao, ``{The Page curve of
  Hawking radiation from semiclassical geometry},''
  \href{http://dx.doi.org/10.1007/JHEP03(2020)149}{{\em JHEP} {\bfseries 03}
  (2020) 149}, \href{http://arxiv.org/abs/1908.10996}{{\ttfamily
  arXiv:1908.10996 [hep-th]}}.

\bibitem{Mathur}
S.~D. Mathur, ``{What is the dual of two entangled CFTs?},''
\href{http://arxiv.org/abs/1402.6378}{{\ttfamily arXiv:1402.6378 [hep-th]}}.

\bibitem{Mahajan}
A.~Almheiri, R.~Mahajan, and J.~Maldacena, ``{Islands outside the horizon},''
\href{http://arxiv.org/abs/1910.11077}{{\ttfamily arXiv:1910.11077 [hep-th]}}.

\bibitem{Sharon}
E.~Sharon and D.~Mumford, ``{2D-Shape Analysis Using Conformal Mapping},''
  \href{http://dx.doi.org/10.1007/s11263-006-6121-z}{{\em Int J Comput Vision}
  {\bfseries 70} (2006) 55--75}.

\bibitem{Cardy}
P.~Calabrese and J.~L. Cardy, ``{Entanglement entropy and quantum field
  theory},'' \href{http://dx.doi.org/10.1088/1742-5468/2004/06/P06002}{{\em J.
  Stat. Mech.} {\bfseries 0406} (2004) P06002},
  \href{http://arxiv.org/abs/hep-th/0405152}{{\ttfamily arXiv:hep-th/0405152}}.

\bibitem{CG}
\url{https://people.eecs.berkeley.edu/~demmel/cs267/lecture24/lecture24.html}.

\end{thebibliography}\endgroup

\end{document}